\documentclass[aps,prb,twocolumn,showpacs]{revtex4}

\usepackage{helvetica}
\usepackage{epsfig}
\usepackage{subfigure}
\usepackage{amssymb}
\usepackage{color} 

\begin{document}

\title{Top quark anomalous couplings at the International Linear Collider}

\author{Erik Devetak}\affiliation{Oxford University, Denys Wilkinson Building, Keble Road, Oxford OX1 3RH, United 
Kingdom}
\author{Andrei Nomerotski}\affiliation{Oxford University, Denys Wilkinson Building, Keble Road, Oxford OX1 3RH, 
United Kingdom}
\author{Michael Peskin \footnote{Work supported by the US Department of Energy under contract DE--AC02--76SF00515.}}  \affiliation{SLAC, Stanford University,
2575 Sand Hill Road, Menlo Park, CA 94025 USA } 
\date{\today}

\begin{abstract}
We present a study of the experimental determination of the forward-backward
asymmetry in the process $e^+e^-\to t\bar t$ and in the subsequent $t\to Wb$ decay, studied in the context of
the International Linear Collider.  This process probes the elementary couplings of the 
top quark to the photon,  the $Z$ and the $W$ bosons at a level of precision that 
is difficult to achieve at hadron colliders.  Measurement of 
the forward-backward asymmetry requires excellent $b$ quark identification
and determination of the quark charge. The study reported here is 
performed in the most challenging all-hadronic channel $e^+e^- \to 
b\bar b q\bar q q\bar q$.  It includes realistic details of the 
experimental environment, a full Monte Carlo simulation of the
detector, based on the Silicon Detector concept, and realistic event 
reconstruction.  The forward-backward asymmetries are determined to a 
precision of approximately 1\% for each of two choices of beam polarization.  We analyze
the implications for the determination of the $t\bar t Z$  and $Wt \bar b$ couplings.


\end{abstract}

\pacs{14.65.Ha}

\maketitle



\section{Introduction}

The top quark is substantially more massive than the other known quarks.
Simply by virtue of this fact, the top quark couples more strongly to 
the particles that generate the spontaneous symmetry breaking of the 
electroweak interactions.  It is possible that the large mass of the
top quark is explained by new interactions of the top quark.  It is 
thus important to measure the properties of the top quark carefully, 
searching for signals of special interactions of this quark.

Particularly interesting quantities to study are the form factors that
describe the coupling of the top quark to elementary currents.  These are
the analogues for any new interactions of the proton form factors, which
played such a large role in the elucidation of QCD.   We will study
the process  $e^+e^-\to t\bar t$.  In this reaction, two sets of form 
factors enter, the form factors that describe the $\gamma$ and $Z$ couplings
to $t\bar t$, which describe the $t\bar t$ production vertex, and the
the form factors that describe the $W$ coupling to $t\bar b$, which 
describe the $t$ and $\bar t$ decay vertices.  As a matter of 
principle, a full reconstruction of the $t\bar t$ system in $e^+e^-$
annihilation can give information on both sets of vertices.  The
effects of the possible form factors on observables of the $t\bar t$
system have been studied by many authors, 
for example, \cite{Kane,Schmidt,Hioki,wtb,Djouadi,Nagano,Bernreuther}.  Some of these couplings
will be constrained at the Large Hadron Collider  (LHC), but others 
are very difficult to access there.  In particular, the vector and 
axial vector couplings of the top quark to the $Z$ boson are 
shifted by new physics effects in many models~\cite{Simmons,Agashe,Berger}.
However,
these couplings are very difficult to measure precisely at the LHC, 
and the associated form factors
are completely inaccessible at values of $Q^2$ larger than $m_Z^2$.

In this paper, we will 	begin a study of the determination of these 
form factors under realistic experimental conditions at the proposed
future $e^+e^-$ collider, the International Linear Collider (ILC).
We will make use of the detector model given by the Silicon Detector
(SiD) concept and the set of full-simulation tools developed for 
the benchmarking of SiD~\cite{SiD}.  These tools provide a very 
detailed simulation of the experimental environment at the ILC.

We will consider the forward-backward asymmetries both for the $b$
and $\bar b$ quarks and for the $t$ and $\bar t$ quarks.  In each case,
the forward-backward asymmetry is defined as:
\begin{equation}
  \label{eq:afb}
  A_{fb}= 
\frac{\sigma(\theta<90^{\circ})-\sigma(\theta>90^{\circ})}{\sigma(\theta<90^{\circ})+\sigma(\theta>90^{\circ})}
\end{equation}
where $\sigma(\theta<90^{\circ})$ is the cross section of the events in which the
$b$ or $t$ quark has a polar angle of less 
than $90^{\circ}$ in the centre of mass frame of reference. The standard spherical coordinate system convention is 
used to define $\theta$.
This asymmetry measurement is
a complex analysis in a dense multi-jet environment. Typical events
have 6-jet final states.  Flavor-tagging must be done to identify 
the $b$ quark jets and resolve the combinatoric ambiguities.  
Quark charges must be measured to distinguish the $t$ and $\bar t$
decay products.  Detector resolution and acceptance together with
non-ideal efficiency and purity of the reconstruction algorithms could play
a critical role in determining the ultimate sensitivity of the 
measurement and hence its physics reach.  This study addresses these
issues for the first time.  Our conclusion is that, with the 
beam conditions and integrated luminosities that the ILC will 
provide,  a well-designed
detector can overcome these potential problems and realize the  
small measurement uncertainties that were projected in parametric 
studies.

The paper is organized as follows:  Section II gives general parameters
of top quark production at the ILC.  Section  III introduces the SiD
detector concept.  Section IV presents the software framework used 
in this analysis.  Section V describes the signal selection and the 
calculation of the cross section for the fully hadronic $t\bar t$
final state.  This section also discusses the flavor-tagging method and
its performance.  Section VI is devoted to the quark charge reconstruction
algorithms, which are fundamental to the analysis.    The results for
forward-backward asymmetries are presented in Section VII.  Section VIII
puts these results in  context by interpreting them as bounds on 
deviations of the $Zt \bar t$ and $Wtb$ form factors from their Standard Model
values.  Section  IX gives out conclusions.

\section{Top Quark at the ILC}

The International Linear Collider is a proposed electron-positron accelerator operating in the centre of mass energy 
range $\sqrt{s} =$ 200 GeV - 500 GeV. An upgrade to the centre of mass energy of 1 TeV is also envisaged as are 
possible calibration runs at the Z boson mass energy \cite{ilc}. The maximum design luminosity is 2 $\times$ 
10$^{34}$ cm$^{-2}$ s$^{-1}$. In the analysis presented here the centre of mass energy and total integrated 
luminosity were assumed to be respectively 500 GeV and 500 fb$^{-1}$, the latter one equivalent to a few years of 
ILC running.

The top quark at the ILC, assuming the 500 GeV operation, is mainly produced in pairs through the  
$e^{+}e^{-}\rightarrow Z \rightarrow t\bar{t}$ and $e^{+}e^{-}\rightarrow \gamma \rightarrow t\bar{t}$ processes. 
The  theoretical total cross-section of top quark pair production is approximately 600 fb \cite{topXsec}. Although 
this value is substantially lower than the one at the Large Hadron Collider (LHC), the clean environment, well 
defined initial state and polarization make the ILC an ideal machine to perform top quark precision measurements. 

\section{The SiD Detector Concept}

\begin{figure}[ht]
\begin{center}
\includegraphics[width=0.45\textwidth]{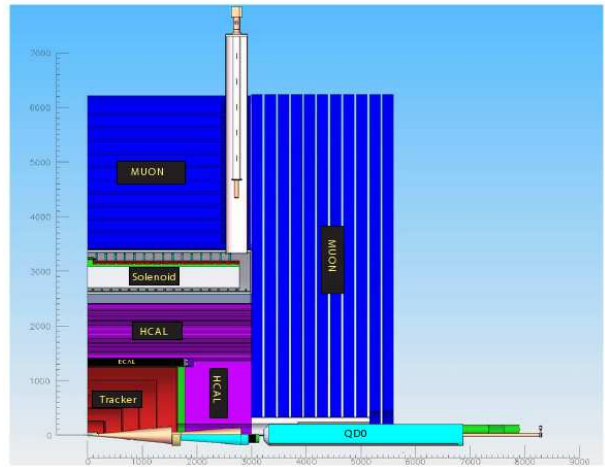}
\caption[Schematic view of the SiD detector]{Disposition of subdetectors in SiD quadrant. All dimensions are in mm.}
\label{f:sidcut}
\end{center}
\end{figure}

The top quark properties are studied with Silicon Detector concept which is a general purpose detector  designed to 
perform precision measurements and at the same time to be sensitive to a wide range of possible new phenomena 
\cite{SiD}. It is based on a five layer silicon pixel vertex detector, silicon tracking with single bunch time 
stamping capabilities, silicon-tungsten electromagnetic calorimetry and a highly segmented hadronic calorimeter. The 
Particle Flow Algorithm (PFA) \cite{PFA} is an important strategy driving the basic philosophy and layout of the 
detector. SiD also incorporates a five Tesla solenoid, an iron flux return and a muon identification system. A 
schematic view of SiD quadrant is shown in Figure \ref{f:sidcut}.

\section{Analysis Framework}

The event generation has been performed using the WHIZARD MC 
generator \cite{WHIZ1,WHIZ2}. Event samples were created with the expected ILC 
baseline parameters of 80\% electron and 30\% positron polarization.  Half 
of the event sample was
created with a positive electron and negative positron polarization, while
the other half has been created with a negative electron and positive 
positron polarization.

WHIZARD was used to generate samples of all 0, 2, 4, and 6 fermion final
states as well as top quark-dominated 8 fermion processes. This generation used electroweak vertices only, with gluon 
emission turned off.  The intent of this strategy was to correctly 
describe multifermion processes such as return to the $Z$ ($e^+e^-\to 
\gamma^* Z^* \to 4$ fermions) and similar processes with intermediate 
off-shell $W$ bosons. These reactions with $t$-channel exchange and 
off-shell electroweak bosons
are the most important backgrounds to multi-fermion $e^+e^-$ annihilation
processes.  QCD was included in the events by using PYTHIA to evolve
final-state quarks through parton showering, fragmentation, and decay.
PYTHIA\cite{pythia} was also used to generate final-state photon radiation.  There
is no double-counting of multi-fermion production between the WHIZARD
stage and the parton shower stage. This procedure treats multi-gluon 
radiation only approximately and ignores quantum interference between the
electroweak and QCD production amplitudes.  However, these
are relatively small effects at the ILC and are unimportant
except in dedicated QCD studies.

About 7 million events were created and processed through the full
GEANT 4\cite{geant4} detector simulation, with individual events weighted to
reflect the statistical sampling. However all of the 6 and 8 fermion states, 
the ones most relevant for the analysis, were left unweighted. The sample has been subsequently divided into 
$b\bar{b}q\bar{q}q\bar{q}$ final states, which constituted the signal and all remaining events representing the 
background. 

In addition to this `pseudo data' events a further independent sample of 2 million  $b\bar{b}f\bar{f}f\bar{f}$ 
events was used for the calibration of algorithms. 

The jet clustering algorithm used in this analysis is the y-cut algorithm  \cite{cluster} with the number of jets 
fixed at six to match the number of jets expected for a hadronic $t\bar{t}$ event.

\section{Top Quark Selection and $e^{+}e^{-}\rightarrow t\bar{t}\rightarrow b\bar{b}q\bar{q}q\bar{q}$ Cross Section}
The analysis starts with a simple event selection based on several global variables described below. Events with 
isolated leptons, defined as a jet containing only one reconstructed particle which is either an electron or a muon 
are rejected as only the $b\bar{b}q\bar{q}q\bar{q}$ final state was considered.

\begin{figure}[ht]
\begin{center}
\subfigure{ \includegraphics[width=0.23\textwidth]{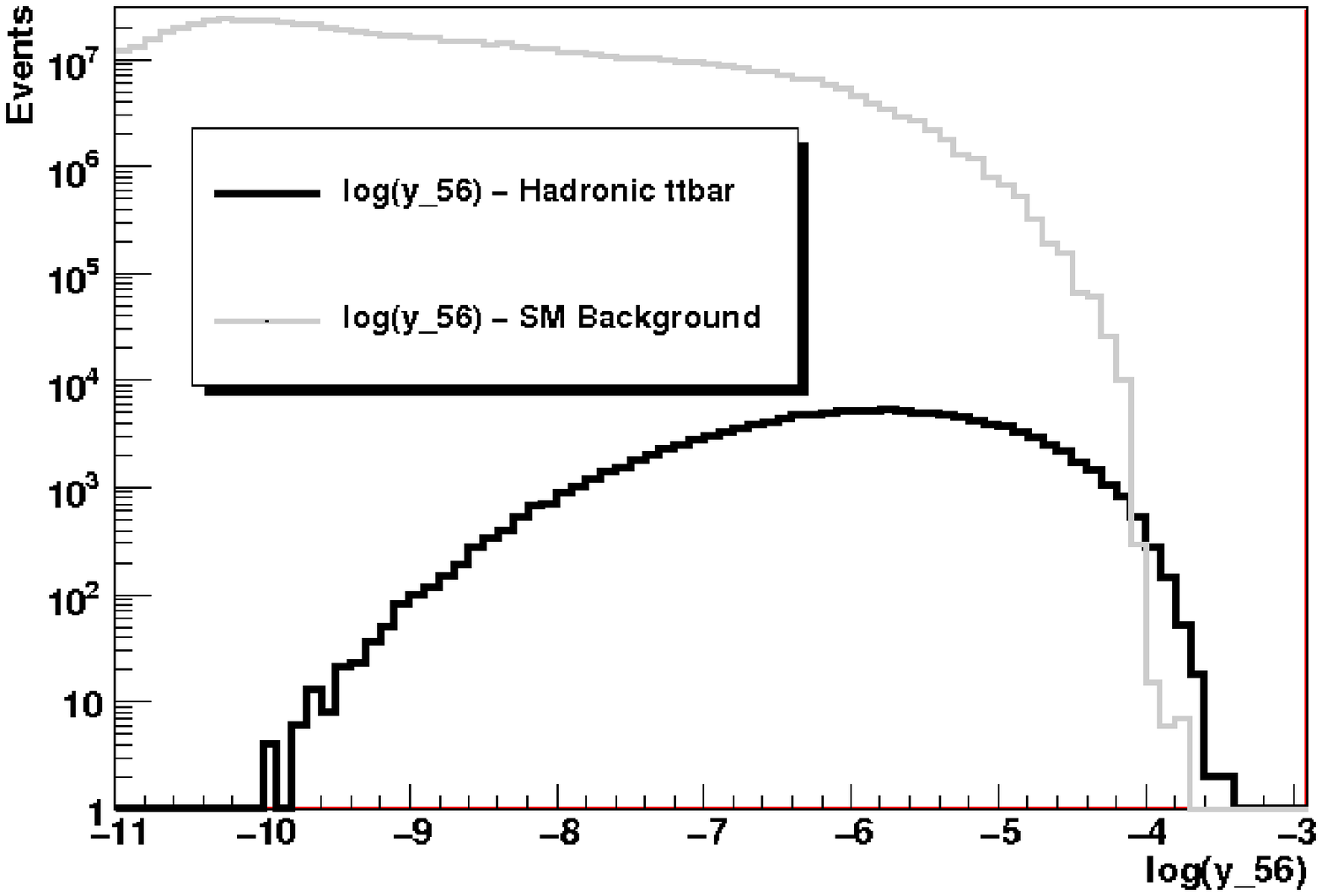}}
\subfigure{ \includegraphics[width=0.23\textwidth]{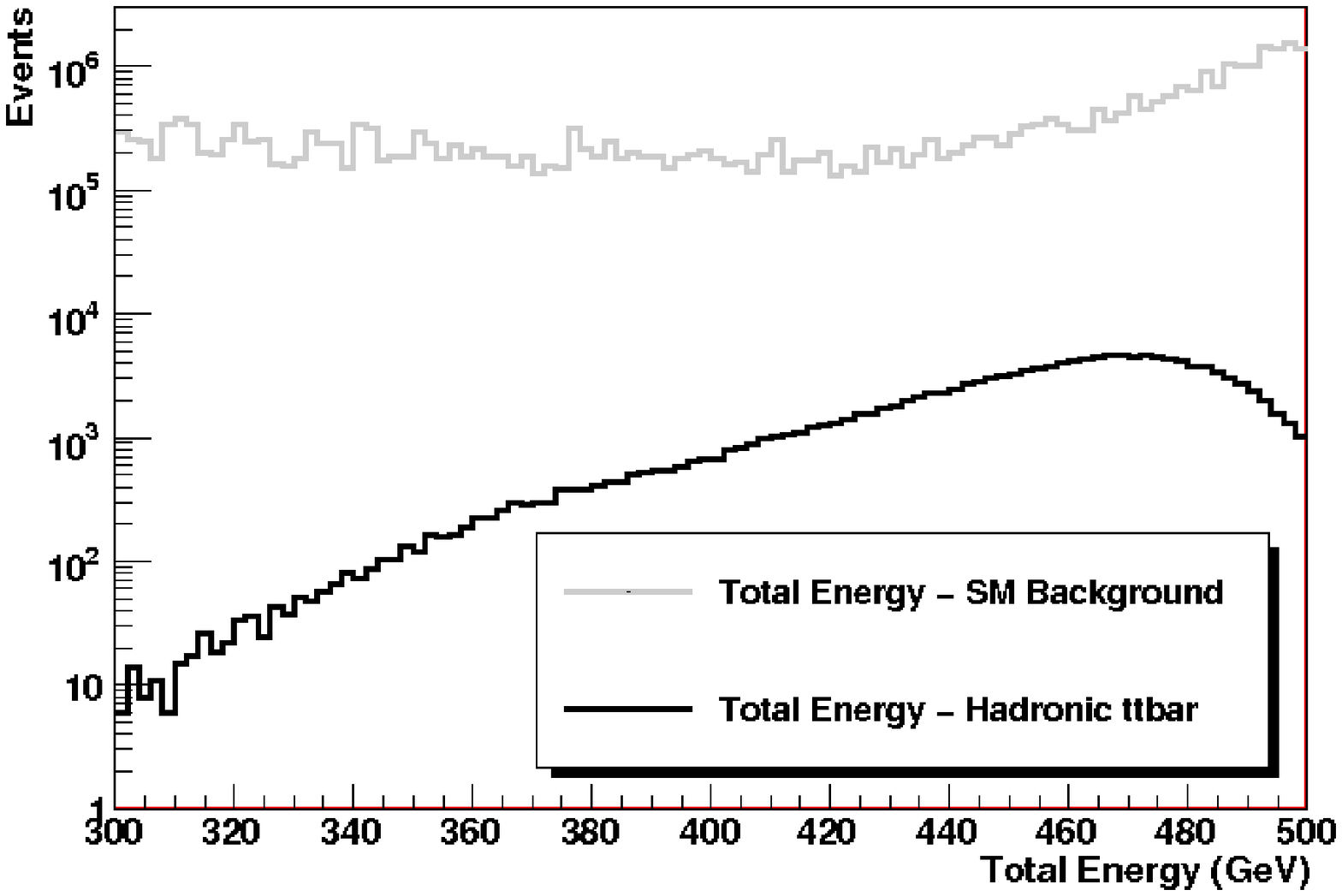}}
\subfigure{ \includegraphics[width=0.23\textwidth]{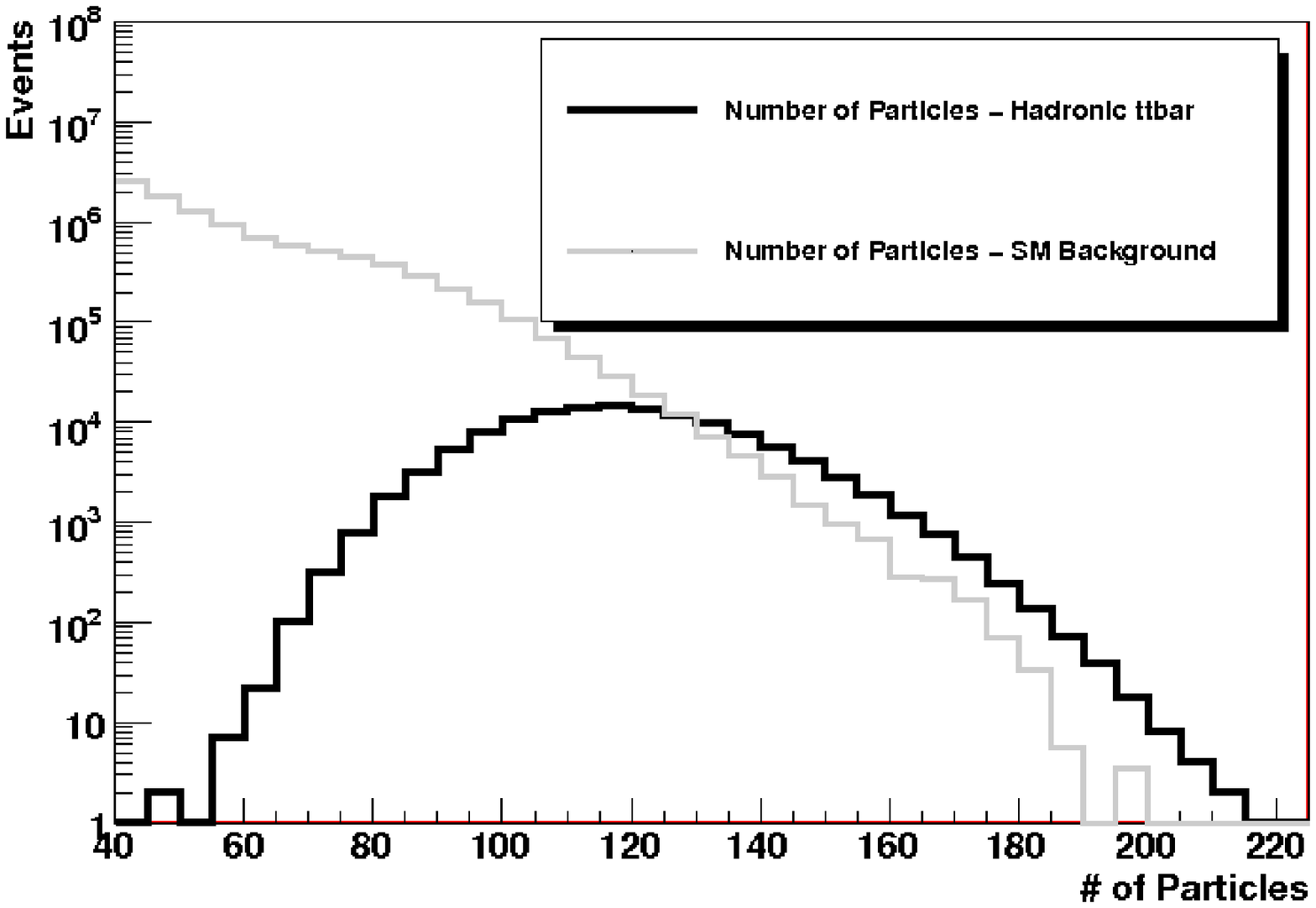}}
\subfigure{ \includegraphics[width=0.23\textwidth]{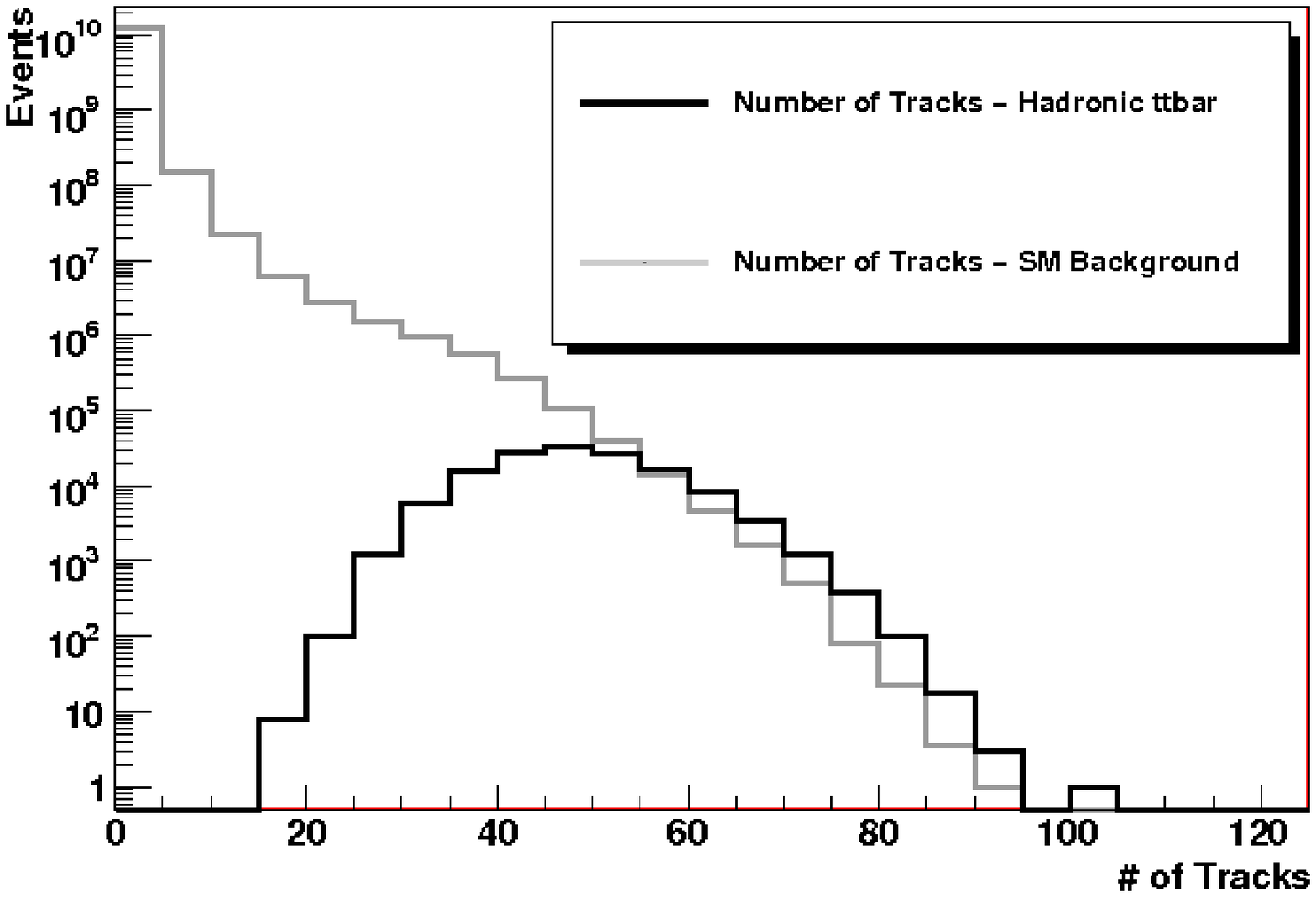}}
\caption[Kinematic and Topological Event Selections] { Kinematic and Topological Event Selections a) $y_{56}$, b) 
total energy, c) number of particles in the event, d) number of tracks in the event.}
\label{f:cuts1}
\end{center}
\end{figure}

Subsequently a set of kinematic and topological discriminating variables has been defined: the total energy of the 
event; the jet finder $y_{56}$ parameter, which represents the y-cut separation between the five and six jet 
hypothesis; the number of particles and the number of tracks. The number of particles in the event is defined as the 
number of reconstructed particles identified by the PFA. Figure \ref{f:cuts1} shows distributions of these variables 
for the signal and background samples before any selections.

Table \ref{t:Eventselect1} presents  the kinematic and topological event selections that have been used.
\begin{table}[ht]
\small
\begin{center}
\begin{tabular}{|c|c|c|} \hline\hline
E$_{tot}$  & $>$ & 400 GeV \\
\hline
log($y_{56}$)  & $>$ & -8.5 \\
\hline
number of particles in event  & $>$ & 80 \\
\hline
number of tracks in event& $>$ & 30 \\
\hline
\hline
\end{tabular}
\normalsize
\caption[Kinematic and Topological Event Selections]{List of the kinematic and topological event selections.}
\label{t:Eventselect1}
\end{center}
\end{table}
After this stage all but 492000 background events have been rejected. This compares to the initial number of 12.5 
$\times$ 10$^{9}$ events. The efficiency loss for the initial 143000 signal events due to this procedure is equal to 
9.7\%. The subsequent stage of the analysis aims to identify the $b$ quarks and to identify the W bosons exploiting 
its significant invariant mass.

\begin{figure}[ht]
\begin{center}
\includegraphics[width=0.45\textwidth]{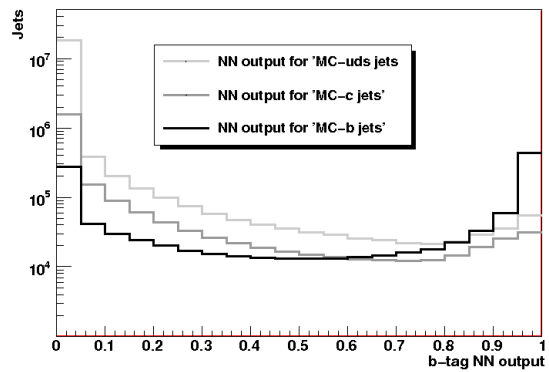}
\caption[Distribution] {Distribution of the flavour tagging neural network output for $uds$, $c$ and $b$ quark 
jets.}
\label{f:bsixjets}
\end{center}
\end{figure}

For the purpose of $b$ quark identification the output of LCFI flavour tagging algorithm \cite{LCFI} has been used 
with the default settings. Figure \ref{f:bsixjets} shows the performance of the LCFI b-tagging algorithm when used 
for a  $e^{+}e^{-}\rightarrow t\bar{t}\rightarrow b\bar{b}q\bar{q}q\bar{q}$ sample. The neural network output for 
$uds$, $c$ and $b$ quark jets demonstrates a good separation of different quark flavours for multi-jet environment. 
In numerical terms, a selection corresponding to the $b$ quark tagging efficiency of 45.0\% will tag 2.6\% of charm 
quarks and 0.8\% of light quarks \cite{thesis}.

\begin{figure}[ht]
\begin{center}
 \includegraphics[width=0.45\textwidth]{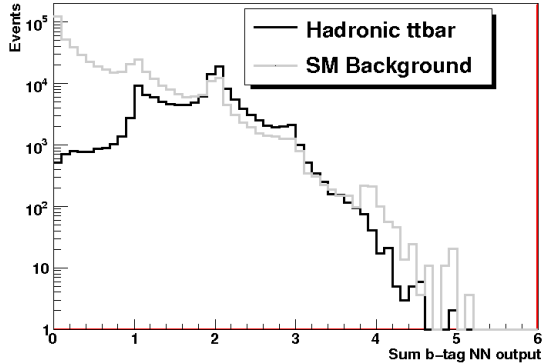}
\caption[Flavour tagging distribution] {The sum of the b-tag neural network outputs of all six jets for the signal 
and the background events after the kinematic and topological event selection.}
\label{f:bsumb}
\end{center}
\end{figure}

Three additional event selections have been applied to the remaining events. The sum of the b-tag neural network 
outputs of all six jets has been required to be higher than 1.5; the b-tag parameter of the most b-like and second 
most b-like jet has been required to be at least 0.9 and 0.4 respectively. Figure \ref{f:bsumb} shows the sum of the 
b-tag of the neural network outputs of all six jets for the signal and background events after the kinematic and 
topological event selection and before any b-tagging selection. It is clear that this is a powerful discriminant to 
select a clean $t\bar{t}$ sample.
In order to identify the invariant mass of the reconstructed W bosons the KinFit  kinematic fitting algorithm 
\cite{kinfit} has been used with a single constraint that the masses of the two W bosons were equal. Only the four 
least b-like jets have been considered for the fit in order to reduce the number of combinations. All the events 
with a W mass of more than 110 GeV or less than 50 GeV have been rejected. 

After the b-quark and W boson identification procedure approximately 74000 $b\bar{b}q\bar{q}q\bar{q}$ signal and 
33500 background events have passed all selections corresponding to signal efficiency of 51.5\% and purity of 
68.8\%. A significant proportion of the remaining background derives from the  $W^{+}W^{-}\rightarrow 
q\bar{q}q\bar{q}$ and $b\bar{b}l\nu q\bar{q}$ events, with a smaller contribution from $ZZ \rightarrow 
q\bar{q}q\bar{q}$. 

The top quark mass has been reconstructed using the same kinematic fitting approach. The primary aim of this 
procedure was to find a correct match of the $b$ quarks to the corresponding W boson, which will be required when 
the polar angle of the top quark needs to be reconstructed later in the analysis. The reconstructed top mass was 
also used to further suppress the background rejecting all events with masses lower than 145 GeV and higher than 195 
GeV. Events that yield a probability of less than 1\% with respect to the constrains used for the fitting are also rejected.  All constraints used for the top mass kinematic fitting can be found in Table \ref{t:topkinconst}. The final efficiency of the wholes selection process is 29.8\% for a purity of 79.7\%.  
\begin{table}[ht]
\small
\begin{center}
\begin{tabular}{|c|c|c|} \hline\hline
Mass(top1) & $=$ & Mass(top2)\\
\hline
Mass(W1) & $=$ & 80.4 GeV\\
\hline
Mass(W2) & $=$ & 80.4 GeV\\
\hline
Mass(b$_{Jet1}$) & $=$ & 5.8 GeV\\
\hline
Mass(b$_{Jet2}$) & $=$ & 5.8 GeV\\
\hline
E$_{Total}$ & $=$ & 500 GeV\\
\hline
p$_{x}$;p$_{y}$;p$_{z}$ & $=$ & 0\\
\hline
\hline
\end{tabular}
\normalsize
\caption[Top Kinematic Fitting Constraints]{List of kinematic fitting constraints used for the calculation of the 
top mass.}
\label{t:topkinconst}
\end{center}
\end{table}

Once the event selection has been performed it is relatively straightforward to calculate the cross section of the 
$e^{+}e^{-}\rightarrow t\bar{t}\rightarrow b\bar{b}q\bar{q}q\bar{q}$ process by the simple use of the following 
formula:
\begin{equation}
  \label{eq:xscalc}
  \sigma= \frac{N_{ALL}-N_{BG}}{\epsilon \int L dt}
\end{equation}
where $N_{ALL}$ is the total number of observed events, while $N_{BG}$ is the number of simulated
background events, $\epsilon$ is the signal selection efficiency and
$\int Ldt$ is the integrated luminosity.Under the assumption that the signal efficiency and the integrated 
luminosity can
be determined with negligible errors and that the background can be reliably determined and subtracted the 
statistical error on the cross section is equal to $\sqrt{N_{ALL}} /(\epsilon \int Ldt)$. The cross section has been 
calculated to be 287.4 $\pm$ 1.3 fb for the whole sample, 370.5 $\pm$ 1.6 fb and 204.3 $\pm$ 1.2 fb for the 
two different polarization samples; the first cross section being for the sample with negative electron 
polarization. It has to be noticed that these are the cross sections for the $e^{+}e^{-}\rightarrow 
t\bar{t}\rightarrow b\bar{b}q\bar{q}q\bar{q}$ process and not for $t\bar{t}$ production. 

\section{Quark Charge Reconstruction}
Next step in the analysis is reconstruction of the quark charge which is necessary to determine the forward backward 
asymmetry of the bottom and top quarks. Hadronization and fragmentation processes obscure the quark charge since the 
bottom quarks fragment into neutral mesons in more than 50\% of the cases. While charged B mesons, when 
reconstructed correctly, allow for unambiguous interpretation of the quark charge, for the neutral B hadrons the 
charge is not representative of the quark charge. Moreover the neutral B mesons oscillate which further dilutes the 
charge reconstruction. 

Several variables sensitive to the charge have been studied and an efficient quark charge estimator has been devised 
as a combination of two variables, the vertex charge and jet charge, as described below. Note that this technique 
considerably improves a simple vertex charge algorithm used in the LCFI Vertex software \cite{LCFI}.

\subsection{Vertex Charge and Jet Charge Algorithms}
\begin{figure}[ht]
\begin{center}
\subfigure{ \includegraphics[width=0.45\textwidth]{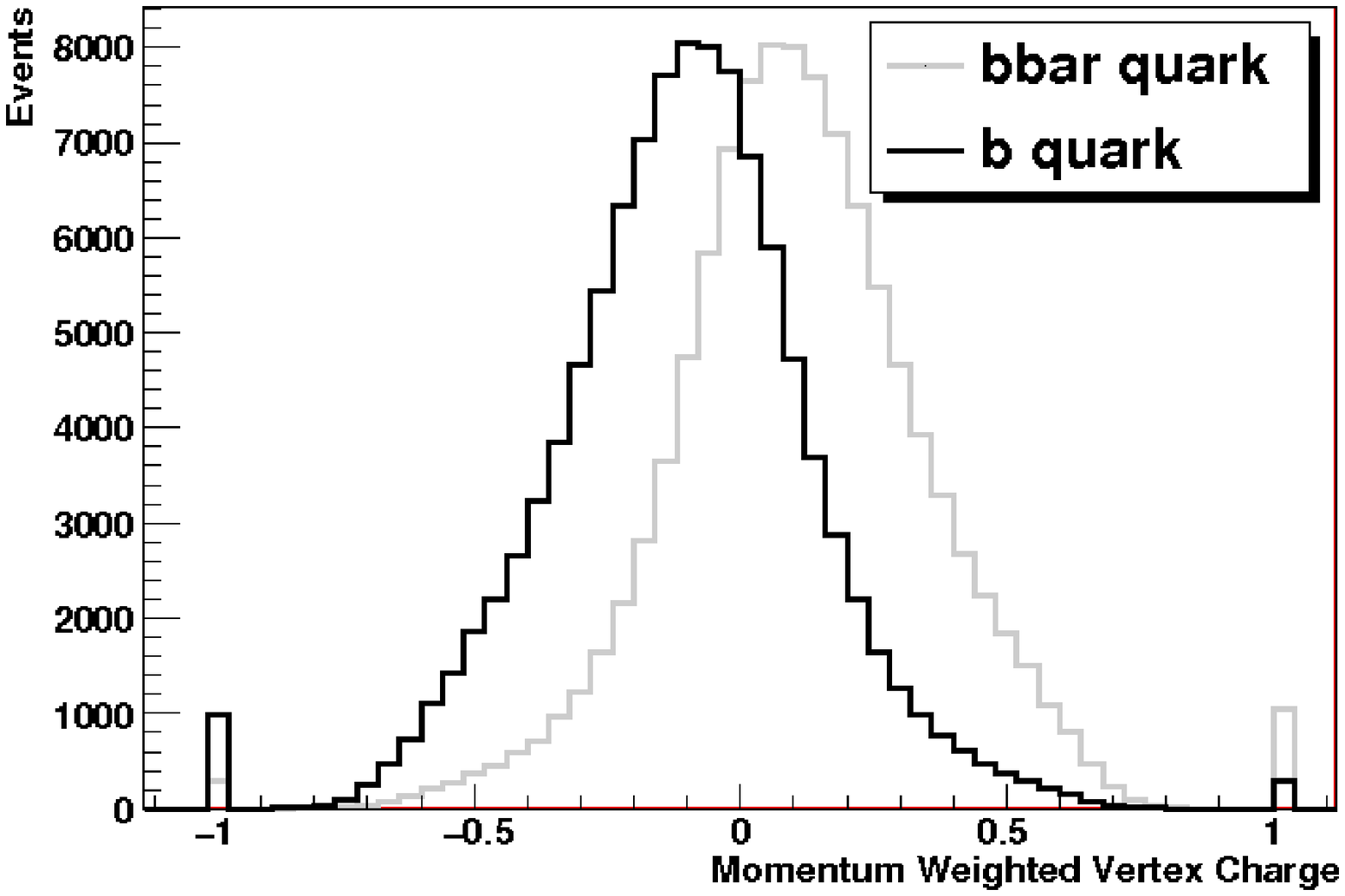}\label{f:lcfimwvcdist}}
\subfigure{ \includegraphics[width=0.45\textwidth]{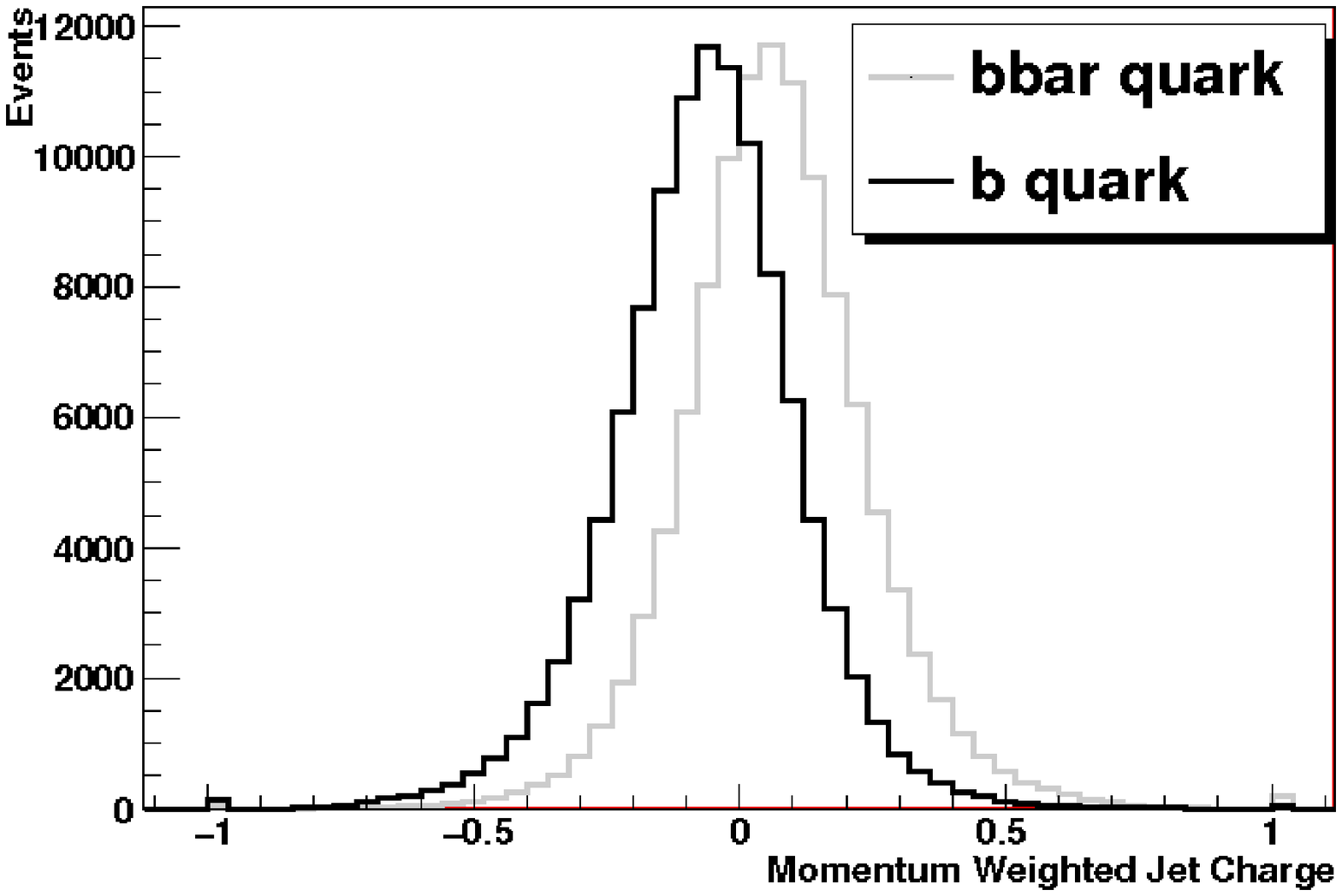}\label{f:lcfimwjcdist}}
\caption[Charge Reconstruction] {Distributions of reconstructed charge for the template signal sample for $b$ quark 
and $\bar{b}$ quark jets a) using the momentum weighted vertex charge,  b) using the momentum weighted jet charge.}
\end{center}
\end{figure}

The vertex charge algorithm uses all tracks associated to a secondary vertex weighted by their momentum to define 
the vertex charge $Q_{VTX}$ as per the following formula:
\begin{equation}
  \label{eq:momweighted}
  Q_{VTX}= \frac{\sum_{j} p_{j}^k Q_{j}}{\sum_{j} p_{j}^k}
\end{equation}
where $Q_{j}$ is the charge of the $j$-th track, $p_{j}$ is the momentum of the track and $k$ is a user defined 
parameter; the sums are performed only on the tracks associated with the vertex. The $k$ parameter was chosen at 0.3 
after optimization. The performance of such method for discriminating the parton charge in the signal sample can be 
seen in Figure \ref{f:lcfimwvcdist}. Only genuine, identified at the MC level $b$ quark jets with a neural net b-tag 
higher than 0.4 were included without any requirement to the $b$ quark final state.

Another method of the quark charge determination implemented in the analysis, momentum weighted jet charge 
\cite{mwvc}, is similar to the one already described in Equation \ref{eq:momweighted} with the only difference in 
the  track selection process which now includes all the tracks present in a jet rather than in a vertex. The jet 
charge algorithm recovers 3.2\% of identified b-jets which do not have a secondary vertex.

The performance of the algorithm can be seen in Figure \ref{f:lcfimwjcdist}. Also in this case the optimal value for 
the $k$ parameter has been determined to be 0.3.

The two algorithms rely on different principles to identify the quark charge. The jet charge algorithm exploits the 
kinematic consideration that the most energetic hadrons have a higher probability of containing the charge of the 
quark that initiated the jet \cite{field}. The principle behind the vertex charge algorithm is instead based on 
precisely determining all the tracks that derive from the displaced vertex due to the $b$ quark considerable 
lifetime. In this case the aim is to directly determine the charge of the meson while the momentum weighting folds 
in information on the reliability of the track.
\begin{figure}[ht]
\begin{center}
\subfigure{ \includegraphics[width=0.45\textwidth]{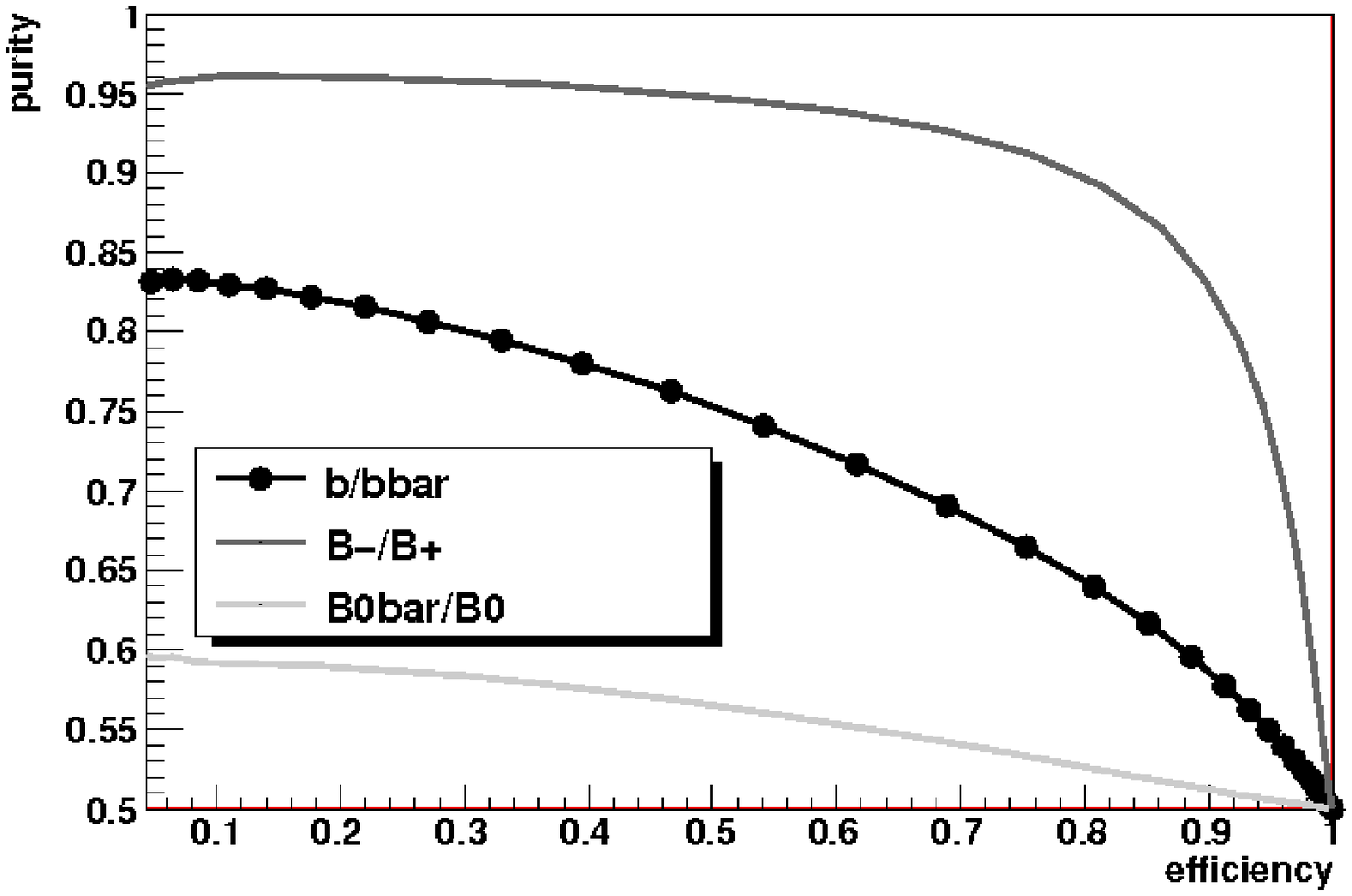}\label{f:bobpmwvc}}
\subfigure{ \includegraphics[width=0.45\textwidth]{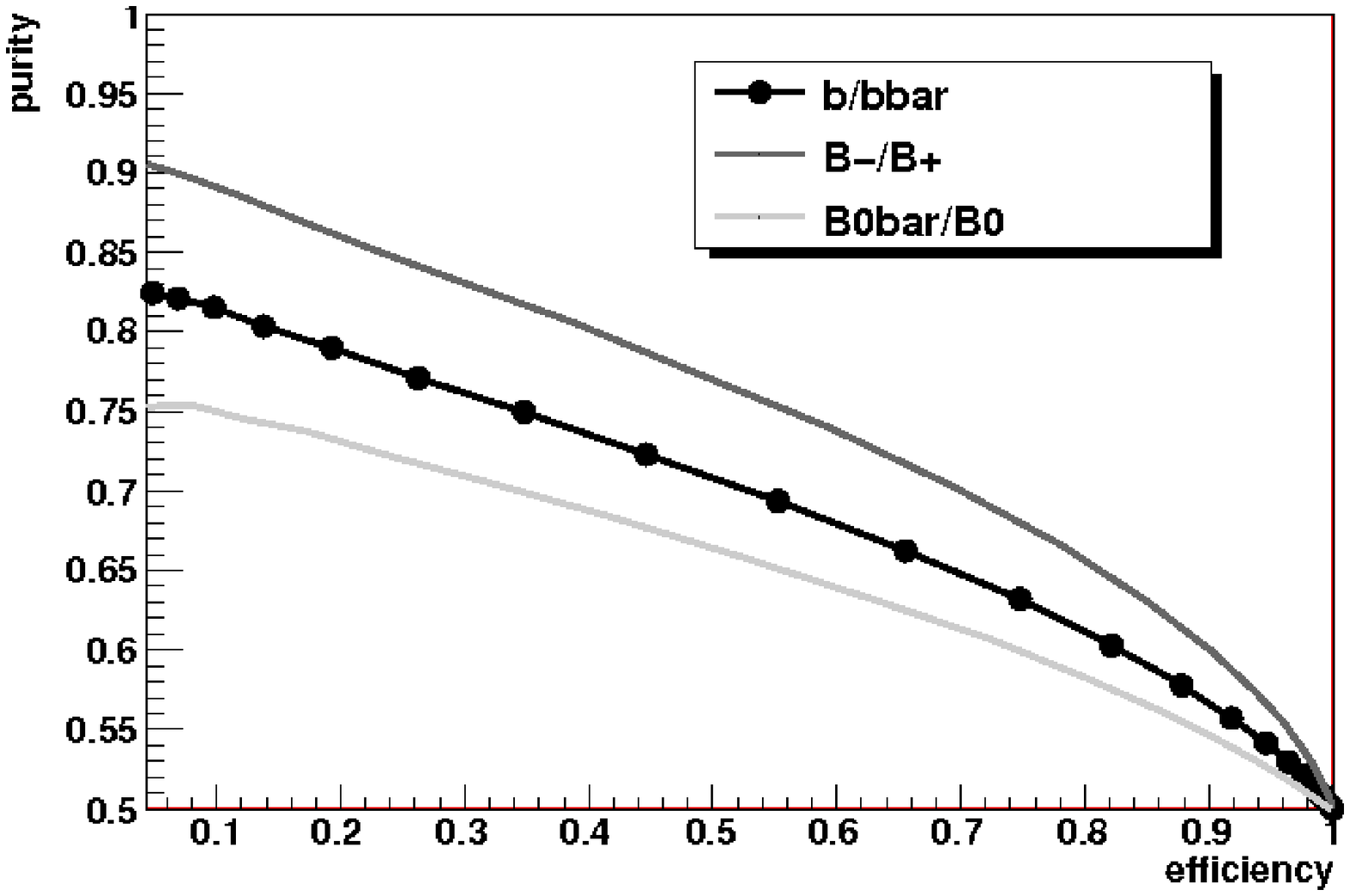}\label{f:bobpmwjc}}
\caption[Momentum Weighted Vertex and Jet Charge: $B^+/B^-$ and $\bar{B^0}/B^0$] { Performance of a) momentum 
weighted vertex charge and b) momentum weighted jet charge in distinguishing $B^+$ from $B^-$, $\bar{B^0}$ from 
$B^0$, and $b$ from $\bar{b}$.}
\end{center}
\end{figure}

It is expected that the vertex charge algorithm is sensitive only to the charged B-mesons while the jet charge 
algorithm is more universal for different B-hadron species. The performance of the algorithms has therefore been 
tested for only charged mesons $B^{+}$ and $B^{-}$ and for only neutral mesons $B^{0}$ and $\bar{B^{0}}$. Figure 
\ref{f:bobpmwvc} shows the purity of a sample with a certain quark charge as a function of selection efficiency. 
This demonstrates that  the momentum weighted vertex charge is able to distinguish well between $B^{+}$ and $B^{-}$, 
while having almost no discriminatory power when it comes to $B^{0}$ and $\bar{B^{0}}$. 

Differently the performance of the momentum weighted jet charge, Figure \ref{f:bobpmwjc}, is more similar between 
the two cases and the algorithm can separate reasonably well also $B^{0}$ and $\bar{B^{0}}$, which included both 
$B^0_d$ and $B^0_s$ mesons. Most of the difference between the charged and neutral mesons in this case can be 
attributed to the flavour oscillations of neutral mesons. While in this process the charge of the meson does not 
change, the charge of the $b$ quark does. This introduces a further dilution in the charge discrimination. The 
effect is rather small  in the $B^0_d$ mesons, which have a period of oscillation larger than their mean lifetime. 
In the case of $B^0_s$ mesons the effect is dominant as oscillations are much faster than the meson lifetime. 

\subsection{Combined Charge}
As the two different methods rely on different information and are rather independent, they have been combined  into 
a single discriminant, based on the probability ratios \cite{d0}. If $f_{i}^{b}(x_i)$ is the probability density 
function for the $b$ quark for variable $x_i$ and $f_{i}^{\bar{b}}(x_i)$ is the equivalent distribution for the 
$\bar{b}$ quark then for each discriminating variable $x_i$ their ratio $r_i$ is defined as:
\begin{equation}
  \label{eq:singlecombinedcharge}
  r_i(x_i)= \frac{f_{i}^{\bar{b}}(x_i)}{f_{i}^{b}(x_i)}
\end{equation}
where the index $i$ denotes the discriminating variable. Distributions of $f^{b}$ and $f^{\bar{b}}$ were determined 
using independent samples.

\begin{figure}[ht]
\begin{center}
\subfigure{ \includegraphics[width=0.45\textwidth]{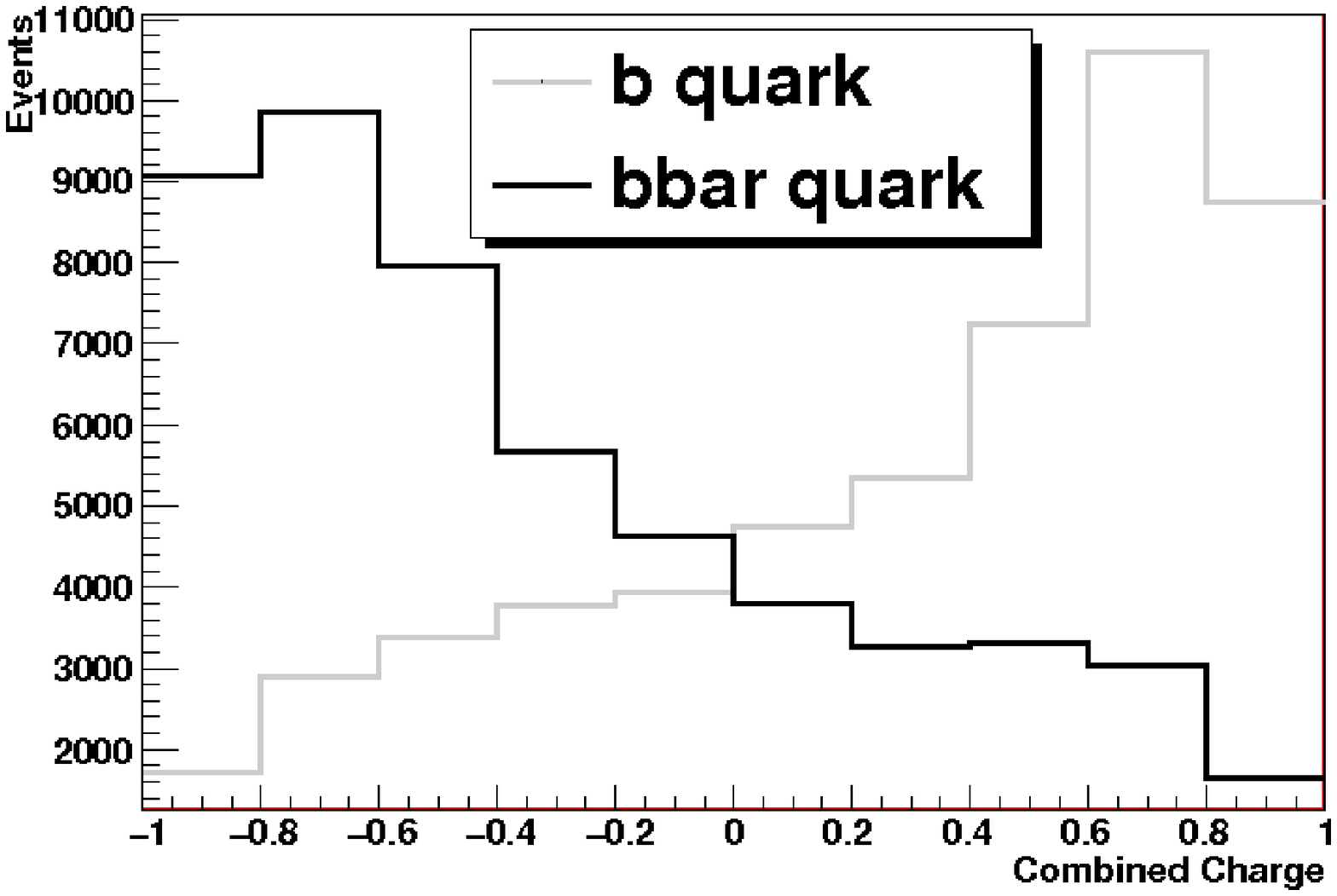} \label{f:combinedcharge}}
\subfigure{ \includegraphics[width=0.45\textwidth]{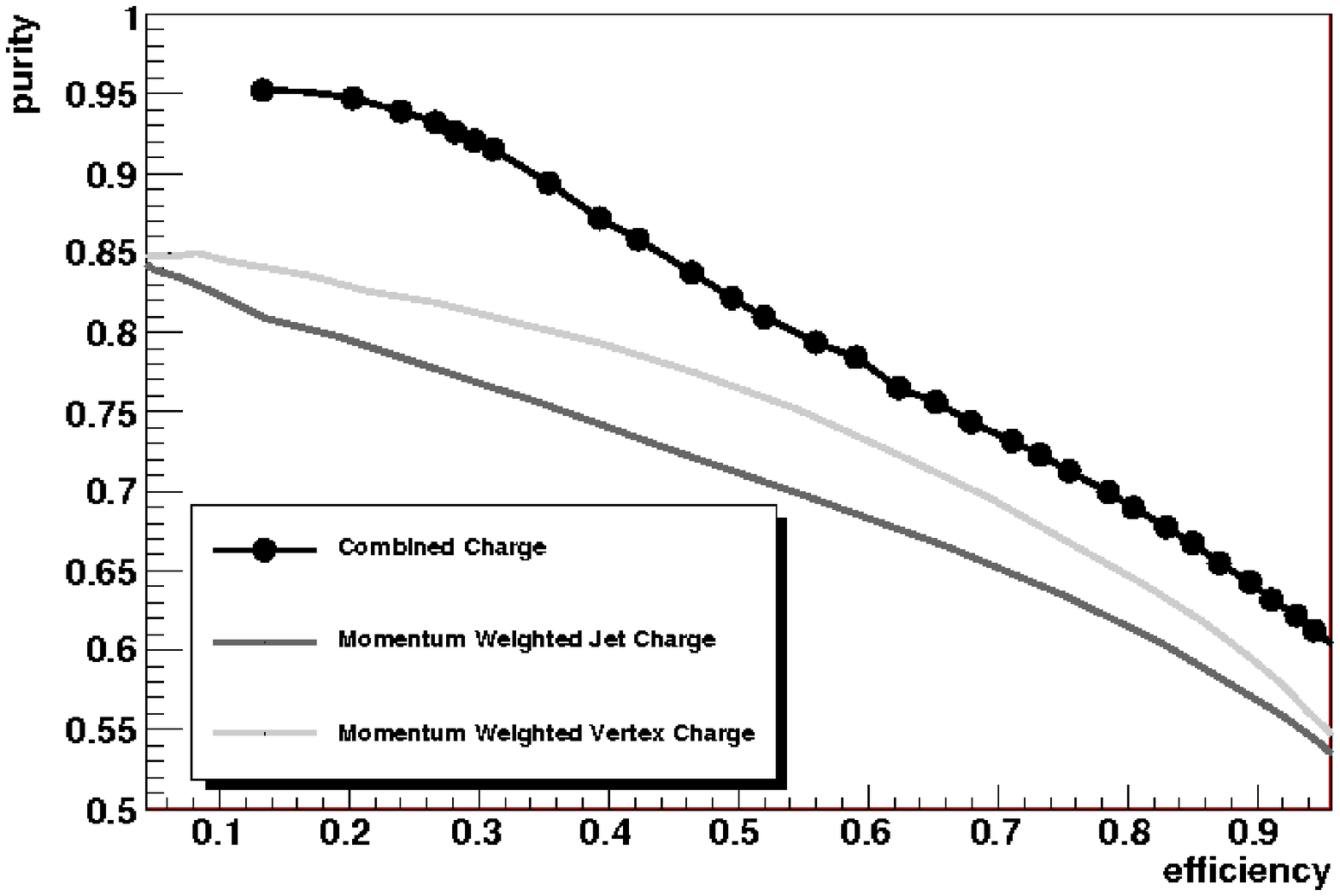}\label{f:combinedchargepe}}
\caption[Combined Charge]{Combined charge a) distributions for $b$ quark and $\bar{b}$ quark jets b) purity versus 
efficiency curves for $b$ quark and $\bar{b}$ quark jets for combined charge, momentum weighted vertex charge and 
momentum weighted jet charge. Shown for the 174.0 GeV sample after all event selections have been applied.}
\end{center}
\end{figure}

For each data event a combined tagging variable can then be defined:
\begin{equation}
  \label{eq:combinedtag}
  r= \prod_i r_{i}
\end{equation}
The range of possible values for $r$ is between 0 and $\infty$. Given the definition of $r$, if $r<1$  then the 
reconstructed jet is more likely to be from a $b$ quark and if $r>1$ the jet is more likely to originate from 
$\bar{b}$ quark. For convenience a variable $C$ changing between -1 and +1 has been defined as: 
\begin{equation}
  \label{eq:combinedcharge}
  C = \frac{1-r}{1+r}.
\end{equation}
A jet with $C>0$ is more likely to derive from a $b$ quark and a jet with $C<0$ is more likely to derive from a 
$\bar{b}$ quark. Figure \ref{f:combinedcharge} shows the combined quark charge performance for the 174.0 GeV top 
quark sample after all event selections have been applied. Figure \ref{f:combinedchargepe} instead shows the purity 
versus efficiency curves for the combined charge algorithm in the same sample when compared to standalone momentum 
weighted vertex charge and momentum weighted jet charge algorithms, as it can be seen in fig. 
\ref{f:combinedchargepe}. The algorithm efficiency is improved by 4\% to 10\% for a purity range from 60\% to 80\%.

The method described above allows a straightforward inclusion of other quark charge discriminants such as  the 
lepton charge \cite{d0} and dipole charge \cite{dipolecharge}.

\section{Quark Forward Backward Asymmetries}

\subsection{Bottom Quark Forward Backward Asymmetry}
Before calculating the forward backward asymmetries for the $b$ and the $t$ quark, the possibility of performing an 
event selection based on the reconstructed charge of the quarks has been investigated. For this purpose one would 
like to use the information derived from both jets. Assuming that the event is actually a 
$b\bar{b}q\bar{q}q\bar{q}$, rather than an event from the SM background, and that the quark identification has been 
correctly performed, the charge calculations performed on the two b-jets are really two uncorrelated measurements of 
the same quantity. The two b-jets must, in fact, have opposite absolute values for their charge. 

\begin{figure}[ht]
\begin{center}
 \includegraphics[width=0.45\textwidth]{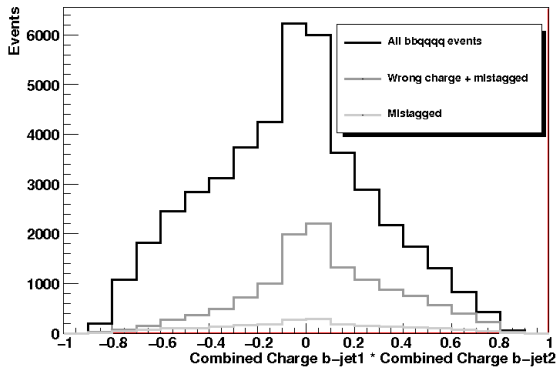}
\caption[Combined Charge b-jet1 $\times$ Combined Charge b-jet2] {(Combined charge b-jet1) $\times$ (Combined charge 
b-jet2) distribution for reconstructed events, mistagged evens and events with misidentified charge. Only 
$b\bar{b}q\bar{q}q\bar{q}$ signal events are used.}
\label{f:ccdist}
\end{center}
\end{figure}

The combined charges of the two jets with the highest neural net b-tags are therefore multiplied and used as an 
event selection parameter. Figure \ref{f:ccdist} shows such distribution for the signal events with explicit 
contributions from mis-identified events where the mis-identification occurred either in the b-tagging (mistagged) 
or in the quark charge determination (wrong charge). The main aim of this procedure would be not to reject the SM 
background but rather to suppress the events in which the $b$ quark has been mistagged or the charge of such quark 
has been misreconstructed. An event charge is labeled as misreconstructed when the reconstructed combined charge of 
the $\bar{b}$ quark jet is higher than the combined charge of the $b$ quark jet.

An optimization has been attempted and the value of $S/\sqrt{S+B}$ has been maximized where $S$ is the number of 
signal events and $B$ is the number of background events including both the SM background and mistagged events. 
Interestingly enough the optimization suggested that all events should be included. Under these conditions the total 
signal efficiency is 22.7\%, while the signal purity is 58.1\%. The impurities derive 45.9\% from the SM background, 
45.0\% from the charge misreconstruction and 9.1\% from the $b$ quark misidentification.

\begin{figure}[ht]
\begin{center}
\includegraphics[width=0.45\textwidth]{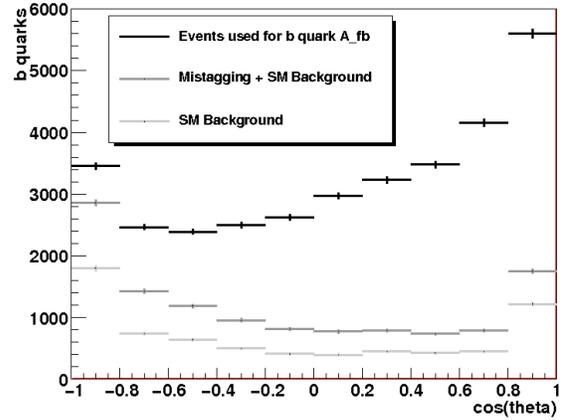}
\caption[Events for the Calculation of the $b$ Quark $A_{fb}$]{Number of events used for the calculation of the $b$ 
quark $A_{fb}$ as function of the $b$ quark $\theta$ angle. In order to qualify as a $b$ rather than $\bar{b}$ 
quark, the combined charge of the jet must be higher than the one of the other b jet present in the event. The 
mistagging refers to both quark charge and flavour.}
\label{f:bafb}
\end{center}
\end{figure}

The calculation of the forward backward asymmetry as defined in Equation \ref{eq:afb} can now be performed using two 
jets with the highest values of neural net b-tags. The jet with a higher combined charge has been declared as 
originating from a $b$ quark, while the other b jet has been declared as originating from a $\bar{b}$ quark. The 
angle $\theta$ of the reconstructed $b$ jet has been used as an approximation the original $b$ quark angle. Figure 
\ref{f:bafb} shows the event distribution with respect to $\cos(\theta)$ of the signal and background events after 
all selections. The mistagged and SM backgrounds peak in the forward regions where the asymmetry is maximal. This 
emphasizes importance of the forward region in the detector design considerations. Note that the mistagged events in 
the distribution include a contribution from $\bar{b}$ quarks which peaks at $\theta=-1$. This explains relatively 
high mistagging rate at $\theta=-1$ when compared to the number of $b$ quarks from the signal. It is because of this 
reason that in the $\theta=1$  region the purity exceeds 60\%, while in the $\theta=-1$ region it is only 15\%. 

The $A_{fb}$ calculation proceeds as follows. The number of correctly reconstructed $b\bar{b}q\bar{q}q\bar{q}$ 
events is evaluated for the forward and backward hemispheres independently. For this the SM background is subtracted 
from the total number of reconstructed events. The number of events left is then multiplied by the purity of the 
reconstruction, accounting for all the events where the charge has been misidentified or where the b-jet has been 
mistagged. The number of correctly identified b-jets is: $N_{b}=(N_{tot}-N_{SM})*p$, where $N_{tot}$ is the total 
number of reconstructed events, $N_{SM}$ is the SM background and $p$ is the purity of the reconstruction. This 
equation is applied to each hemisphere, correspondingly separate purities have been calculated for the forward and 
the backward hemispheres. 
In principle, the number of events should also be corrected for the signal efficiency because Equation \ref{eq:afb}
uses the cross sections. However the efficiencies in the forward and backward regions to a good approximation cancel 
each other out and produce a negligible effect on the final result.This also leads to robustness of the measurement to variations of fragmentation and hadronization models which results in a negligible systematic uncertainty.
Table \ref{t:bass} shows the $A_{fb}$ results for different event selections. The first line corresponds to the case 
of no selection which maximizes the sensitivity as discussed above\cite{D0topcharge,Abazov:2011ck,Heister:2004wr,Aaltonen:2011kc}.

\begin{table}[ht]
\small
\begin{center}
\begin{tabular}{|c|c|c|c|c|} \hline\hline
\emph{Event Selection} & \emph{$A_{fb}$}& \emph{$\sigma_{1}$}& \emph{$\sigma_{2}$}& \emph{$\sigma_{3}$}\\
\hline
Charge b$_1$ $\times$ Charge b$_2$ $<$ 1.0 & 0.293 & 0.006 & 0.007 & 0.008 \\
\hline
Charge b$_1$ $\times$ Charge b$_2$ $<$ 0.5 & 0.293 & 0.006 & 0.007 & 0.008 \\
\hline
Charge b$_1$ $\times$ Charge b$_2$ $<$ 0.0 & 0.289 & 0.007 & 0.008 & 0.009 \\
\hline\hline
\end{tabular}
\normalsize
\caption[Reconstructed $A_{fb}$ for the $b$ quark]{Reconstructed $A_{fb}$ for the $b$ quark and the respective 
uncertainties. The different uncertainties ($\sigma_{1}$,$\sigma_{2}$,$\sigma_{3}$) have been calculated with 
different assumptions as explained in the text.}
\label{t:bass}
\end{center}
\end{table}

For each event selection the uncertainty has been calculated with three different assumptions. The lowest 
uncertainty, $\sigma_{1}$, assumes that the efficiency of tagging and the standard model background have been 
perfectly simulated at the MC level and therefore do not contribute to the uncertainty of the forward backward 
asymmetry. The only uncertainty contribution therefore is $\sqrt{N_{tot,\theta <(>) 90^{\circ}}}$ where 
$N_{tot,\theta <(>) 90^{\circ}} $ is the total number of events with $b$ quarks reconstructed in the forward 
(backward) region of the detector. For the second evaluation, $\sigma_{2}$, the statistical uncertainty from the b-tagging efficiency is added in quadrature to the value of $\sigma_{1}$. The added statistical uncertainty is calculated from the previously mentioned ad-hock generaded calibration sample. Finally the third evaluation, $\sigma_{3}$, considers also an additional contribution 
from the statistics of background samples \cite{thesis}, which is added in quadrature to $\sigma_{2}$. In each of the three cases the uncertainty has been 
calculated separately for the forward and backward regions and subsequently the standard error propagation has been 
used to evaluate the $A_{fb}$ uncertainty.

The calculated asymmetry agrees well with initial asymmetry at the MC level, 0.291, which suggests that the 
performed analysis has not introduced any systematic bias. In order to check for any significant detector smearing 
leading to systematic effects in $A_{fb}$ the angular resolution of the $b$ jet $\theta$ angle with respect to the 
original $b$ quark has been determined. The resolution has been found to be 0.08 radians, and therefore its effect 
on the reconstructed asymmetry is negligible. 

Finally the result can also be decomposed with respect to the different polarizations used. In the case of -80\% 
electron polarization and +30\% positron polarization the asymmetry has been calculated to be 0.356 with an 
uncertainty of 0.010. In the case +80\% electron polarization and -30\% positron polarization the asymmetry has been 
calculated to be 0.155 with an uncertainty of 0.012. In both cases the 
$\sigma_{3}$ definition of error is being used. 

\subsection{Top Quark Forward Backward Asymmetry}
The analysis of the top quark asymmetry is similar to the one already presented for the $b$ quark. The only added 
complication is that, differently from the $b$ quark, where the angle $\theta$ of the $b$ jet can be used as a very 
good approximation to the angle $\theta$ of the original $b$ quark, the direction of the top quark must be 
reconstructed from its decay products, using the kinematic fitter to determine correct pairing of two $b$ quarks and 
two W bosons. More specifically the direction of the top quark is calculated from the combination of jets that 
minimizes the $\chi^2$ of the fit given the constraints stated in Table \ref{t:topkinconst}. 
\begin{figure}[ht]
\begin{center}
\includegraphics[width=0.45\textwidth]{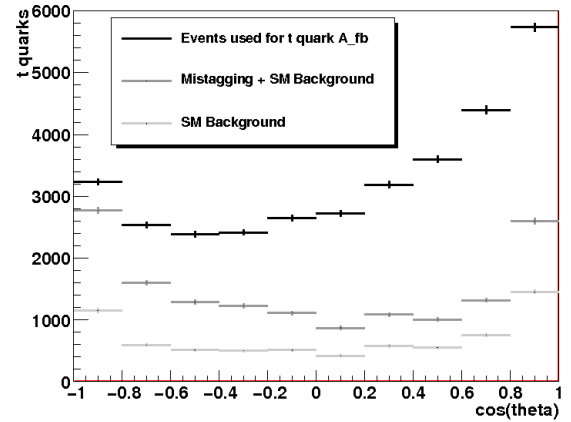}
\caption[Events for the Calculation of the $t$ Quark $A_{fb}$]{Number of events used for the calculation of the top 
quark $A_{fb}$ as function of the top quark $\theta$ angle. In order to qualify as a $t$ rather than $\bar{t}$ 
quark, the combined charge of the b-jet used to reconstruct the top quark must be higher than the charge of the 
other b-jet present in the event.}
\label{f:tafb}
\end{center}
\end{figure}

The charge of the top quark is determined through the charge of the daughter $b$ quark. If a reconstructed $b$ quark 
jet is part of the three jets used to reconstruct the top quark then the top quark is declared as a $t$. If instead 
a $\bar{b}$ jet is present the quark is declared as a $\bar{t}$. Given the constrains set to the kinematic fitter 
only one such quark will be present in each jet. Figure \ref{f:tafb} shows the distribution of top quark events with 
respect to their $\cos(\theta)$. The distribution includes the SM and  mistagging backgrounds. 

Subsequently the same ${A_{fb}}$ calculations have been performed as the ones described in the previous section for 
the $b$ quark case with results shown in Table \ref{t:tass}. The calculated asymmetry agrees well with the initial 
one at the MC level, 0.351.

\begin{table}
\begin{center}
\begin{tabular}{|c|c|c|c|c|} \hline\hline
\emph{Event Selection} & \emph{$A_{fb}$}& \emph{$\sigma_{1}$}& \emph{$\sigma_{2}$}& \emph{$\sigma_{3}$}\\
\hline
Charge b$_1$ $\times$ Charge b$_2$ $<$ 1.0 & 0.356 & 0.006 & 0.007 & 0.008 \\
\hline
Charge b$_1$ $\times$ Charge b$_2$ $<$ 0.5 & 0.348 & 0.006 & 0.007 & 0.008 \\
\hline
Charge b$_1$ $\times$ Charge b$_2$ $<$ 0.0 & 0.353 & 0.007 & 0.008 & 0.009 \\
\hline\hline
\end{tabular}
\normalsize
\caption[Reconstructed $A_{fb}$ for the $t$ Quark]{Reconstructed $A_{fb}$ for the $t$ quark and the respective 
uncertainties. The different uncertainties ($\sigma_{1}$,$\sigma_{2}$,$\sigma_{3}$) have been calculated with 
different assumptions as explained in the text.}
\label{t:tass}
\end{center}
\end{table}

Finally, in the same fashion as for the $b$ quark, the $\theta$ angle resolution has been found equal to 0.19 
radians. This will have a negligible contribution to the total calculated $A_{fb}$ because only the very central 
events of the Figure \ref{f:tafb} distribution will ever be smeared enough to change hemisphere when reconstructed. 
The asymmetry in this region is however small and does not affect the total $A_{fb}$. 

Similary to the asymmetry of the bottom quark the achievable statistical precision has been calculated also for the 
two cases of polarized beams. Unsurprisingly values identical to the ones presented for the bottom quark (0.010 and 
0.012) have been found.

\section{Discussion}
To put these results in context, we will interpret them in terms of 
constraints on the couplings of the top quark to the vector bosons, $t\bar t Z$ and $Wtb$.  

As we have pointed out already in the Introduction, many models of new
physics predict large corrections to the left- and right-handed 
vector $t\bar t Z$ couplings.  The measurement of forward-backward
asymmetries in $e^+e^- \to t\bar t$ will allow these couplings to be
determined experimentally in a very direct way.

As a starting point for the analysis, we define the $\gamma$ and $Z$
vertex form factors of the top quark by
\begin{eqnarray}
\label{eq:Pesk1}
 \nonumber {\cal L}^Z_{t\bar{t}} &=&  e A_\mu \bar t [\gamma^\mu (P_L F_{L\gamma} + P_R F_{R\gamma})
        + i {\sigma^{\mu\nu}q_\nu\over 2 m_t} F_{2\gamma} ] t 
\\& & +  e Z_\mu \bar t [\gamma^\mu (P_L F_{L Z} + P_R F_{R Z})
	  + i {\sigma^{\mu\nu}q_\nu\over 2 m_t} F_{2Z} ] t  
\end{eqnarray} 
 where $P_L$ and $P_R$ are the left and right handed chiral projection operators,  $q_\nu$ is the 4-momentum of the virtual photon or $Z^0$ and 
$\sigma^{\mu\nu}=i/2(\gamma^{\mu}\gamma^{\nu}-\gamma^{\nu}\gamma^{\mu})$.
The tree-level Standard Model values of the form factors are
\begin{eqnarray}
\nonumber & & F_{L\gamma} = F_{R\gamma} = {2\over 3}  \qquad  
F_{2\gamma} = F_{2Z} = 0 
\\& &  F_{LZ} = {({1\over 2}-{2\over 3}s_w^2)\over s_wc_w} \quad
  F_{RZ} = {(-{2\over 3}s_w^2)\over s_wc_w} 
\label{Fvals}
\end{eqnarray}
where $s_{w}$ and $c_{w}$ are the sine and cosine of the Weinberg angle. 
For later reference, the numerical values of the Standard 
Model $Z$ boson form factors, using $s_w^2 = 0.231$, are
$$  F_{LZ} = 0.821 \qquad   F_{RZ} = - 0.365  \ .  $$
In principle, we could also introduce in each line a fourth, CP-violating,
form factor proportional to $\sigma^{\mu\nu}\gamma^5$.  One might
also include contact interactions between the $e^+e^-$ and $t\bar t$
states~\cite{Hiokitwo}.  

In principle, a complete helicity analysis of the full set of 
production and decay angles has the power to constrain many of these 
parameters independently.  However, in this paper, we have concentrated
on the experimental measurement of the forward-backward asymmetries.
Since our main concern here is to illustrate the power of that measurement,
we will choose a parametrization with two free parameters that can be 
determined in terms of the two top quark forward-backward asymmetries corresponding 
to two cases of beam polarizations.

In the following, then, we will assume that the $\gamma t\bar t$ form
factors take their Standard Model values given in (\ref{Fvals}), that the
magnetic moment $Z$ form factor $F_{2Z}$ is zero, and that the decay 
form factors take their Standard Model values. We will
allow only values of the $Zt\bar t$ form factors $F_{LZ}$ and $F_{RZ}$ to be
varied, and we will determine these parameters from two measurements of the 
$t\bar t$ forward-backward asymmetry with different beam conditions.
The choice of -80\% electron polarization and +30\% positron polarization
leads to $t\bar t$ production dominantly from the initial state $e^-_Le^+_R$.
In the Standard Model, for this polarization choice, the $\gamma$ and $Z$
$s$-channel amplitudes interfere constructively for the production of 
$t_L\bar t_R$ and destructively for the production of $t_R\bar t_L$, 
leading to a large positive forward-backward asymmetry.  The main effect 
of changing the $Z$ form factors is to relax the destructive interference
in the production of $t_R\bar t_L$.  Thus, the asymmetry in this polarization
state is mainly sensitive to $F_{RZ}$, which gives the larger effect on 
the $t_R\bar t_L$ state.  Similarly, the 
choice of +80\% electron polarization and -30\% positron polarization
leads to $t\bar t$ production dominantly from the initial state $e^-_Re^+_L$.
In the Standard Model, for this polarization choice, the $\gamma$ and $Z$
$s$-channel amplitudes interfere constructively for the production of 
$t_R\bar t_L$ and destructively for the production of $t_L\bar t_R$.  This 
also leads to a large positive  forward-backward asymmetry, but one that 
is mainly sensitive to $F_{LZ}$.  Thus, the measurement of the $t\bar t$
forward-backward asymmetry with these two beam settings sensitively 
picks out non-Standard contributions to the two separate $Zt\bar t$
vector form factors~\cite{Barklow}.

For 100\% polarized beams, the sensitivity of the $t\bar t$ forward-backward
asymmetries to deviations of the $Z$ form factors from their Standard Model
values can be computed to be
\begin{equation}
 \pmatrix{\delta A_{FB}(LR) \cr \delta A_{FB}(RL)\cr } 
 \pmatrix{0.138 &  -0.392\cr 0.461 & -0.106 \cr}  
\pmatrix{\delta F_{LZ} \cr \delta F_{RZ} \cr } \
\label{beforetans}
\end{equation}
using
 $\sqrt{s} = 500$ GeV and $\sin^2\theta_w = 0.231$. 
The large off-diagonal terms in this matrix show clearly the effect 
discussed in the previous paragraph.  For an electron 
polarization of -80\% and a positron polarization of +30\%, the fraction of
events in the two relevant initial polarization states is
\begin{eqnarray}
\nonumber  f(e^-_Le^+_R) &=& {(1+ P(e^-))(1+P(e^+))\over 4} = 0.585  
  \\f(e^-_Re^+_L) &=& {(1- P(e^-))(1-P(e^+))\over 4} = 0.035  
\end{eqnarray}

By taking this into account, it is possible to transform the matrix presented in Equation  \ref{beforetans} in order to account for 
the beam polarizations actually used.  Recomputing the numerator and denominator for $A_{FB}$, we find that the relation between
the form factor deviations becomes
\begin{equation}
 \pmatrix{\delta A_{FB}(LR) \cr \delta A_{FB}(RL)\cr } 
 \pmatrix{0.164 &  -0.374\cr 0.367 & -0.238 \cr}  
\pmatrix{\delta F_{LZ} \cr \delta F_{RZ} \cr } \ .
\end{equation}
Then the standard uncertainties reported in Section VII,
\begin{equation}
    \sigma( A_{FB}(LR)) = 0.010   \qquad  \sigma( A_{FB}(RL)) = 0.012 
\end{equation}
gives the uncertainties on  $\delta F_{LZ}$ and $\delta F_{RZ}$,
\begin{equation}
   \sigma(\delta F_{LZ}) = 0.051  \qquad  \sigma(\delta F_{RZ} ) = 0.042 
\end{equation}
with some correlation between the values.  Normalizing to the Standard 
Model values of these parameters,
\begin{equation}
       \sigma(\delta F_{LZ})/F_{LZ} = 0.062   
\qquad  \sigma(\delta F_{RZ} )/F_{RZ} = 0.116  
\end{equation}
These uncertainties are comparable to the values suggested in 
\cite{Barklow} on the basis of parametric simulations.  One can see,
for example, by comparing the models discussed in \cite{Berger}, that 
such measurements would cut deeply into the space of deviations 
predicted in models of new physics. 

The ILC study of the reaction $e^+e^- \to t\bar t$ will also include events in which either the $t$ or the $\bar t$ decays leptonically. These events add a data set of approximately equal size to the one considered here in which the $t/\bar t$ charge discrimination is unambiguous.  Thus, the full analysis of the ILC data will do even better at determining the $Z t \bar t$ couplings.

In a similar manner the results can also be interpreted with respect to the  $Wtb$ anomalous couplings.
As a matter of fact, the decay form factors of the top quark are already constrained at the 
20\% level by the measurement of
the $W$ helicity at hadron colliders~\cite{CDF}, and these measurements
will be improved at the LHC.  Thus, it is likely that, by the time
the ILC operates, the decay form factors could be fixed to experimentally
determined values. Nevertheless, for completeness, we consider the effects of these anomalous couplings, following the notation in \cite{wtb}.

In this case the appropriate vertex under consideration is: 
\begin{eqnarray}
  \label{eq:WTBvertex}
    \nonumber L^{W}_{tb} = & & - \frac{g}{\sqrt{2}} [
W_{\mu}^{-} \bar{b}\left(\gamma_{\mu}A_{L}P_{L}+\gamma_{\mu}A_{R}P_{R}\right)t 
   \\ & & -\frac{1}{2M_W}W_{\mu\nu}\bar{b}\sigma^{\mu\nu}\left(B_{L}P_{R}+B_{R}P_{L}\right)t]
\end{eqnarray}
where $W_{\mu\nu} = D_\mu W_\nu - D_\nu W_\mu$ , $D_\mu = \partial_\mu - ieA_\mu$. $A_{L,R}$ and $B_{L,R}$ are the 
coupling form factors. In the Standard Model  A$_L$ is equal to one, while all the other form factors are equal to 
zero.


Table \ref{t:bassprediction} presents predictions of the $b$ quark asymmetry for different values of the $Wtb$ 
anomalous couplings \cite{wtb}.
\begin{table}[ht]
\small
\begin{center}
\begin{tabular}{|c|c|c|} \hline\hline
\emph{$B_{R}$} & \emph{$B_{L}$} & \emph{$A_{fb}$}\\
\hline
0.0 & 0.0 & 0.279 \\
\hline
0.0 & -0.2 & 0.243 \\
\hline
0.0 & -0.4 & 0.218 \\
\hline
0.0 & -0.6 & 0.197 \\
\hline
0.0 & -1.0 & 0.169 \\
\hline-0.6 & 0.0 & 0.301 \\\hline-1.0 & 0.0 & 0.315 \\\hline\hline\end{tabular}\normalsize
\caption {$A_{fb}$ asymmetry of $b$ quark from the top decay for the Standard Model and anomalous $Wtb$ vertices. 
Calculated at a centre of mass energy of 500 GeV and in the centre of mass rest frame.}
\label{t:bassprediction}
\end{center}
\end{table}

It can be inferred from Table  \ref{t:bassprediction} that the measurement of the the b quark forward-backward asymmetry is sensitive
to the presence of a $B_L$ anomalous form factor whose absolute value is greater than approximately 0.05.  Measurement
of other observables, not considered in this paper,  that specifically target the top quark decay properties will put much stronger constraints on both $B_L$ and $B_R$.

Note that there is a difference between the asymmetry predicted by the Whizard generator which was used for these 
studies, 0.291, and the asymmetry by the CompHEP MC generator \cite{Bcomphep,Pcomphep} used when calculating the 
theoretical predictions \cite{wtb}, 0.279. Part of the discrepancy can also be explained by the fact that the 
generated signal sample is an all inclusive $e^{+}e^{-} \rightarrow \bar{b}q\bar{q}q\bar{q}$ rather than $e^{+}e^{-} 
\rightarrow t\bar{t} \rightarrow b\bar{b}q\bar{q}q\bar{q}$ as assumed in the theoretical paper. In any event, this 
difference is not significant for the purpose of sensitivity estimation.

\section{Conclusion}
We therefore conclude that the achievable resolution for the forward backward asymmetry of the top quark at the ILC 
in the $e^{+}e^{-}\rightarrow t\bar{t}\rightarrow b\bar{b}q\bar{q}q\bar{q}$ channel is approximately 0.008 for a 
total luminosity of 500 fb$^{-1}$. Similarly the achievable resolution for the $b$ quark resulting from the top 
decay is also 0.008. In the case of polarized beams the achievable resolution for both the top and bottom quark 
asymmetries is 0.010 and 0.012 for the -80\% electron polarization, +30\% positron polarization and the +80\% 
electron polarization, -30\% positron polarization respectively. This result allows to constrain the theoretically 
predicted deviations from the Standard Model in the presence of an anomalous coupling of the $Zt\bar{t}$ and $Wtb$ 
vertices. In the case of the $Zt\bar{t}$ coupling the resolution on the predicted Standard Model form factor is of 
the order of 0.05 and 0.04 for the $F_{LZ}$ and the $F_{RZ}$ couplings respectively. In the case of the $Wtb$ the 
performed analysis is sensitive to the presence of an $B_{L}$ anomalous form factor greater that approximately 
0.05.
 The analysis employed realistic detector simulations and advanced reconstruction algorithms in the framework of the 
Silicon Detector concept. A new quark charge reconstruction algorithm used to discriminate between bottom quarks and 
their anti-quarks allowed to achieve a selection purity of up to 80\% for an efficiency of about 60\%.

\textbf{ACKNOWLEDGMENTS}

We would like to thank the colleagues from the SiD software and benchmarking groups, in particular Jan Strube, Tim 
Barklow, Norman Graf and John Jaros for assistance with sample processing and useful discussions. The work of Michael Peskin is supported by the US Department of Energy under contract DE--AC02--76SF00515.

\end{document}